\documentclass[]{spie}  %>>> use for US letter paper
\usepackage{amsmath,amsfonts,amssymb}
\usepackage{graphicx}
\usepackage[colorlinks=true, allcolors=blue]{hyperref}
\usepackage{comment}
\usepackage{booktabs}
\usepackage{adjustbox}

\title{Optimization of a CCD-in-CMOS TDI detector's operating clock voltages by Taguchi based Grey relational analysis}

\author[a]{Swaraj Bandhu Mahato}
\author[a]{Pierre Boulenc}
\affil[a]{imec, Kapeldreef 75,B-3001 Leuven, Belgium}

\authorinfo{Further author information: (Send correspondence to Swaraj Bandhu Mahato)\\Swaraj Bandhu Mahato: E-mail: swaraj.mahato@imec.be}

\begin{document} 
\maketitle

\begin{abstract}

In recent years, CCD-in-CMOS TDI image sensors are becoming increasingly popular for many small satellite missions to assure a fast and affordable access to space for Low Earth Observation. Our monolithic CCD-in-CMOS TDI imager features a specifically developed technology which combines the benefits of a classical CCD TDI with the advantages of CMOS System-On-a-Chip (SoC) design. Like CCD, this detector is also controlled by a large number of clock voltages. Optimizing these voltages allows to increase the performance of the detector by improving multiple characteristic parameters, such as full well capacity (FWC), dark current, linearity, dark signal non-uniformity (DSNU) and charge transfer efficiency (CTE). Traditionally, it has been the standard practice to adjust the CCD voltages by trial and error methods to get a better image. Because of the large parameter space, such subjective procedures may yield far from the optimum performance. This paper reports a design of experiments (DOE) technique applied on the clock voltages to improve the multiple performance parameters of the detector. This method utilizes the Taguchi's orthogonal arrays of experiments to reduce the number of experiments with different voltage combinations. Finally, optimal combination of clock voltages is obtained by converting the multiple performance parameters of the detector into a single Grey relational grade. In this process, the sequences of obtaining parameter values are categorized according to the performance characteristics. The condition Higher-the-better is used for parameters like FWC and CTE whereas condition Lower-the-better is applied for parameters, such as dark current, linearity error and DSNU.

\end{abstract}


\section{INTRODUCTION}
\label{sec:intro}  % \label{} allows reference to this section

Monolithic CCD-in-CMOS image sensors are becoming increasingly popular for many small satellite missions as it combines the noiseless charge transfer of CCD with the efficiency of CMOS high-speed drivers and readout for Time Delay Integration (TDI). This work features our TDI imager features a specifically developed CCD-in-CMOS technology \cite{DeMoor14,Bello14,Boulenc17,Swaraj19}. This CCD-in-CMOS platform was realized by adding a few process modules to a standard $0.13 \mu m$ CMOS process flow containing dual gate oxide nMOS and pMOS transistors (1.2 and 3.3V) enabling monolithic integration of CCD row drivers and fast 12-bit analog-to-digital converters (ADCs) at each column. In order to achieve multi-spectral imaging, this sensor has 7-band with 256 TDI stages each, using 4096 columns at a pixel pitch of $5.4 \mu m$. As each band uses individual on-chip sequencers and CCD drivers for the four-phase CCD pixels, the 7 bands can be configured independently. Backside processing and Anti-Reflective Coating (ARC) have been optimized to enhance quantum efficiency (QE) with 96 \% peak QE at 510 nm wavelength for visible tuned ARC and 93 \% peak QE at 310 nm wavelength for UV tuned ARC. 

Each CCD is characterized by a number of electro-optical properties, which must meet a set of specifications in order to reach the redundancy. Unlike QE, a number of properties can be affected by the modification of the operating voltages used to control the read-out and charge transfers in the CCD pixel. In this optimization work, we are focusing on the clock voltages applied to control the CCD phase operations. Optimizing these voltages allows to increase the performance of the sensor by improving multiple characteristic parameters. Traditionally, it has been the standard practice to adjust the CCD voltages by trial and error methods to get a better image \cite{Janesick01}. For large parameter space, such subjective procedures may yield far from the optimum performance. Because of the number of operating voltages and their possible inter-dependency, an individual performance parameter of a sensor will depend on many of the applied voltages, and adjustments to a single voltage will affect multiple parameters.  The purpose of this study was to find a methodology to fine-tune the controlling voltages to reach optimum CCD performance in a large parameter space. This paper reports a design of experiments (DOE) technique applied on the clock voltages to improve the multiple performance parameters of the CCD detector.  This method utilizes the Taguchi's orthogonal arrays of experiments to reduce the number of experiments with different voltage combinations. Finally, optimal combination of clock voltages is obtained by converting the multiple performance parameters of the detector into a single Grey relational grade.

 \section{CCD Parameters and  design of experiments (DOE) technique}
\begin{figure}[ht]
\centering
\includegraphics[width=0.8\linewidth]{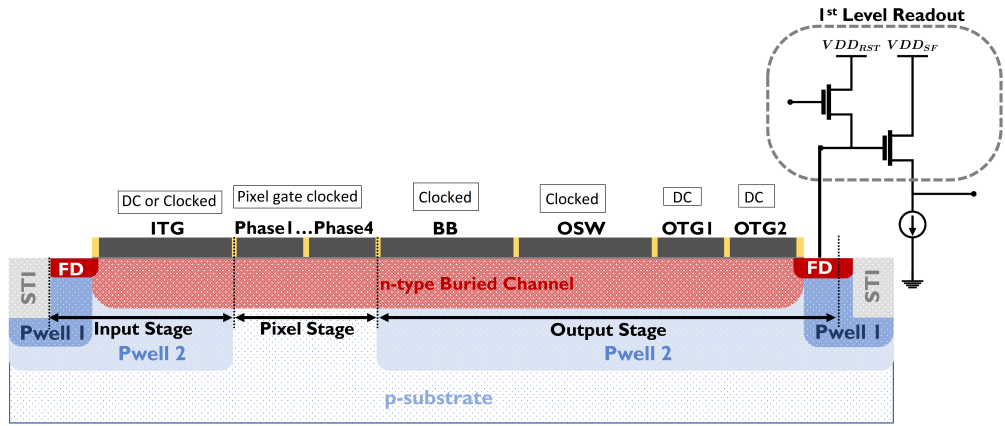}
\caption{CCD-in-CMOS column schematic.}
\label{fig:CCD-in-CMOS}
\end{figure}

 A simple schematic of our 4-phase CCD-in-CMOS column is shown in Figure \ref{fig:CCD-in-CMOS} where phase1...phase4 refer to the four serial clock transfer gates per pixel. In the output stage, the output summing well (OSW) functions as a reservoir for the signal from gate electrodes. A back barrier gate (BB) is needed to prevent the charges from spilling back into the CCD column, when they are drained into the floating diffusion. Output transfer gates (OTG1 and OTG2) serve as static barriers for the OSW during signal accumulation. They are set to a fixed voltage above the lower clock signal level of the OSW, such that the charges will overflow to the floating diffusion, once OSW is driven low. In this paper multiple characteristic parameters, such as full well capacity (FWC), dark current (DC), linearity error (LE), dark signal non-uniformity (DSNU) and charge transfer efficiency (CTE) are studied in the context of CCD voltage optimization.

Through the experiment, we found that the controlling voltages such as Serial Clock High (VDDH\_CCD), Serial Clock Low (VDDL\_CCD), Back Barrier Clock Low (VSSL\_BB), Output Summing Well Clock Low (VSSL\_OSW), Output transfer gate (VSSL\_OTG) have obvious affect on different performance parameters mentioned above. Table \ref{tab:volatage_label} presents the six voltages and their levels designed in the experiments. 
\\
\begin{table}[ht]
    \centering
    \begin{tabular}{|c|c|c|c|c|}
     \hline
        Symbol & Voltages & Level 1 & Level 2 & Level 3\\
        \hline
        A & Serial Clock High ($VDDH\_CCD$) & 1.2V & 1.5V & 1.75V \\
        \hline
        B & Serial Clock Low ($VDDL\_CCD$) & -1.75V & -1.5V & -1.25V \\
        \hline
        C & Back Barrier Clock Low $(VSSL\_BB) = VDDL\_CCD-0.1$ & -1.85V & -1.6V & -1.35V \\
        \hline
        D & Output Summing Well Clock Low $(VSSL\_OSW) = VDDL\_CCD$ & -1.75V & -1.5V & -1.25V \\
        \hline
        E & 1st Output transfer gate $(VSSL\_OTG1) = const. DC$ & -0.7V & -0.7V & -0.7V \\
        \hline
        F & 2nd Output transfer gate $(VSSL\_OTG2) = const. DC$ & 0V & 0V & 0V \\
    \hline    
    \end{tabular}
    \caption{Experimental levels of different voltages.}
    \label{tab:volatage_label}
\end{table}

\section{Taguchi method and results}
Taguchi method is a systematic and efficient approach to find the optimum combination of input parameters. This method utilizes the orthogonal array of experiments. The Taguchi experimental design of $L_9$ orthogonal array has been selected for this study. We only considered $VDDH\_CCD$ and $VDDL\_CCD$ for this orthogonal array as $VSSL\_BB$ and  $VSSL\_OSW$ are dependent on these voltages as mentioned in Table \ref{tab:volatage_label} and output transfer gates ($VSSL\_OTG1$ and $VSSL\_OTG2$) are set to fixed DC voltages above the lower clock signal level of the OSW. Our Taguchi experimental design of L9 orthogonal array shown in Table \ref{tab:Taguchi_label}.

\begin{table}[ht]
    \centering
    \begin{tabular}{|c|c c|}
     \hline
        Trial number & A & B \\
        \hline
        1 & 1.2V & -1.75V \\
        2 & 1.2V & -1.5V \\
        3 & 1.2V & -1.25V \\
        4 & 1.5V & -1.75V \\
        5 & 1.5V & -1.5V \\
        6 & 1.5V & -1.25V \\
        7 & 1.75V & -1.75V \\
        8 & 1.75V & -1.5V \\
        9 & 1.75V & -1.25V \\
    \hline    
    \end{tabular}
    \caption{Taguchi experimental design of L\_9 orthogonal array.}
    \label{tab:Taguchi_label}
\end{table}

Considering the above Taguchi experimental design of L9 orthogonal array complete DOE for all voltages and  the results of experiments were shown in Table \ref{tab:Taguchi_result}.

\begin{table}[ht]
    \centering
    \begin{adjustbox}{width=1\textwidth}
    \begin{tabular}{l cccccc ccccc}
    \toprule
     & \multicolumn{6}{c}{Voltages[V]} & \multicolumn{5}{c}{Characteristic parameters} \\
    \cmidrule(lr){2-7} \cmidrule(lr){8-12}
    Trial & A & B & C & D & E & F & FWC[DN] & DSNU(DN) & LE[\%] & DC[e-/s/pix] & CTE[\%] \\
    \midrule
    1 & 1.20 & -1.75 & -1.85 & -1.75 & -0.7 & 0 & 30315 & 76 & 7.15 & 5445 & 99.92825 \\
    2 & 1.20 & -1.50 & -1.60 & -1.50 & -0.7 & 0 & 34882 & 104 & 1.28 & 4839 & 99.99881 \\
    3 & 1.20 & -1.25 & -1.35 & -1.25 & -0.7 & 0 & 36304 & 88 & 2.11 & 5054 & 99.99864 \\
    4 & 1.50 & -1.75 & -1.85 & -1.75 & -0.7 & 0 & 33914 & 52 & 3.12 & 12529 & 99.99729 \\
    5 & 1.50 & -1.50 & -1.60 & -1.50 & -0.7 & 0 & 41545 & 76 & 2.27 & 5465 & 99.99857 \\
    6 & 1.50 & -1.25 & -1.35 & -1.25 & -0.7 & 0 & 42371 & 64 & 2.42 & 5497 & 99.99865 \\
    7 & 1.75 & -1.75 & -1.85 & -1.75 & -0.7 & 0 & 34757 & 79 & 1.72 & 6338 & 99.99844 \\
    8 & 1.75 & -1.50 & -1.60 & -1.50 & -0.7 & 0 & 38460 & 76 & 1.70 & 5865 & 99.99879 \\
    9 & 1.75 & -1.25 & -1.35 & -1.25 & -0.7 & 0 & 39824 & 73 & 1.62 & 6509 & 99.99851 \\
    \bottomrule
    \end{tabular}
    \end{adjustbox}
    \caption{Complete DOE for all voltages and  the results of experiments.}
    \label{tab:Taguchi_result}
\end{table}

\begin{figure}[ht]
\centering
\includegraphics[width=0.8\linewidth]{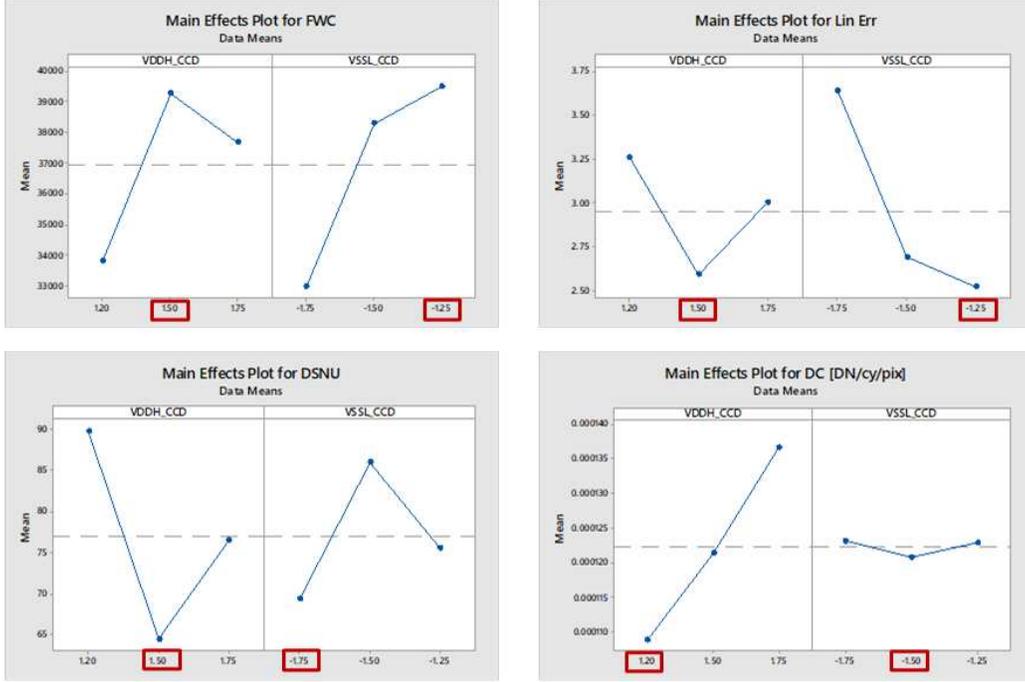}
\caption{Main effect plot for different characteristic parameters.}
\label{fig:DOE_results}
\end{figure}

When looking at the response from the above mentioned DOE, it is not straightforward to find a single optimum voltage combination as also depicted in the Figure \ref{fig:DOE_results}. Here we are looking for the optimum voltages to achieve high FWC, high CTE, low linearity error (Lin Err), low DSNU and low dark current (DC). 

As Taguchi method concentrated on the optimization of single-objective problems. For our large parameter space, it is difficult to compare between the different input factors (CCD voltages) because they exert a different influence. Therefore, the standardized transformation of these factors must be done. 

\section{Optimization methodology using Grey relational analysis (GRA)}
In our optimization work, we used the Grey relational approach along with the Taguchi method. In this method, the multiple performance characteristic can be converted into a single Grey relational grade \cite{Wang06}. All characteristic performance parameters obtained from the Taguchi analysis (see Table \ref{tab:Taguchi_result}) have to be normalized in the range of 0 to 1 to avoid different units and to reduce the variability. It is required since the variation of one response can differ from other response. 
For the higher-the-better parameters like FWC and CTE the original responses are normalized as:

\begin{equation}
    X_i^*(k) = \frac{x_i(k)-min~x_i(k)}{max~x_i(k)-min~x_i(k)}
\label{eq:higherTheBetter}
\end{equation}
where, $i = 1,..., m; k = 1,...,n; m$ is the number of experimental data and n is the number of responses. $X_i^*(k)$ represents the normalized response, $x_i(k)$ represents the original response obtained from the Taguchi analysis, largest value of $x_i(k)$ is $max~x_i(k)$, smallest value of $x_i(k)$ is $min~x_i(k)$. For the lower-the-better parameters like DC, Linearity error, DSNU, the original responses are normalized as:

\begin{equation}
    X_i^*(k) = \frac{max~x_i(k)-x_i(k)}{max~x_i(k)-min~x_i(k)}
\label{eq:lowerTheBetter}
\end{equation}
After completing the normalization for each characteristic performance parameters of our sensor, in order to express a relationship between actual and ideal normalized values, a Grey relational coefficient is determined for each measurement sequence, as expressed in Eq. \ref{eq:GRC}:

\begin{equation}
    \xi_i(k) = \frac{\Delta_{min}+p\Delta_{max}}{\Delta x_i(k)+p\Delta_{max}}
\label{eq:GRC}
\end{equation}
where $\Delta x_i(k)$ is the deviation sequence, which is calculated by, $\Delta x_i(k)=\parallel X_0^*(k)-X_i^*(k)  \parallel$. $\Delta_{min}$ and $\Delta_{max}$ are the minimum and maximum values of the absolute differences ($\Delta x_i(k)$) of all comparing sequences. $p$ is distinguishing or identification coefficient and the range is between 0 to 1. Since our multiple performance characteristic of the CCD consist of both higher-the-better and lower-the-better, $p$ is assumed to be $0.5$ in this case.

After calculating the Grey relational coefficient for each measurement sequence, we calculated the Grey relational grade (GRG) as a weighted sum of the Grey relational coefficients. In our case, we considered equally weighted response. Then, GRG is obtained by averaging the Gray relational coefficients associated with each performance characteristic. It can be expressed as:
\begin{equation}
    \gamma_i=\frac{1}{n}\sum_{k=1}^{n}\xi_i(k)
\label{eq:GRG}
\end{equation}
where, $\gamma_i$ is the required Grey relational grade for $i^{th}$ experiment and $n$ = number of response characteristics. The GRG represents the level of correlation between the reference sequence and the comparability sequence and is the overall representative of all the quality characteristics. Thus our multiple performance parameter optimization problem is converted into single response optimization problem through the Grey relational analysis coupled with Taguchi approach. Then an optimal level of CCD voltages is determined using higher Grey relational grade that indicates the better CCD performance matrix. The Grey relational coefficient, Grey relational grade and the rank of each experiment were found from Table \ref{tab:GRA_result} and the results too. The GRG values offer a single representation for the selected five CCD characteristic performance parameters such as FWC, DSNU, dark current, linearity and CTE and a higher value of GRG is chosen. From Table \ref{tab:GRA_result}, it is found that experiment number 6 has the highest Grey relational grade of 0.851. 
\begin{table}[ht]
    \centering
    \begin{adjustbox}{width=1\textwidth}
    \begin{tabular}{l ccccc ccccc l l}
    \toprule
     & \multicolumn{5}{c}{Normalized Characteristic parameters} & \multicolumn{5}{c}{GRC} \\
    \cmidrule(lr){2-6} \cmidrule(lr){7-11}
    Trial & FWC & DSNU & LE & DC & CTE & FWC & DSNU & LE & DC & CTE & GRG & Rank \\
    \midrule
    1 & 0.0000 & 0.5377 & 0.0000 & 0.9211 & 0.0000 & 0.3333 & 0.5196 & 0.3333 & 0.8638 & 0.3333 & 0.477 & 9 \\
    2 & 0.3787 & 0.0000 & 1.0000 & 1.0000 & 1.0000 & 0.4459 & 0.3333 & 1.0000 & 1.0000 & 1.0000 & 0.756 & 5 \\
    3 & 0.4967 & 0.3047 & 0.8593 & 0.9720 & 0.9976 & 0.4984 & 0.4183 & 0.7805 & 0.9469 & 0.9953 & 0.728 & 6 \\
    4 & 0.2984 & 1.0000 & 0.6863 & 0.0000 & 0.9976 & 0.4161 & 1.0000 & 0.6145 & 0.3333 & 0.9587 & 0.665 & 8 \\
    5 & 0.9314 & 0.5349 & 0.8309 & 0.9186 & 0.9967 & 0.8794 & 0.5181 & 0.7474 & 0.8600 & 0.9934 & 0.800 & 2 \\
    6 & 1.0000 & 0.7682 & 0.8056 & 0.9143 & 0.9978 & 1.0000 & 0.6833 & 0.7201 & 0.8537 & 0.9956 & 0.851 & 1 \\
    7 & 0.3684 & 0.4869 & 0.9254 & 0.8050 & 0.9948 & 0.4419 & 0.4936 & 0.8702 & 0.7194 & 0.9897 & 0.703 & 7 \\
    8 & 0.6755 & 0.5280 & 0.9288 & 0.8665 & 0.9998 & 0.6065 & 0.5144 & 0.8754 & 0.7893 & 0.9996 & 0.757 & 4 \\
    9 & 0.7886 & 0.5944 & 0.9418 & 0.7829 & 0.9958 & 0.7029 & 0.5521 & 0.8959 & 0.6972 & 0.9916 & 0.768 & 3 \\
    \bottomrule
    \end{tabular}
    \end{adjustbox}
    \caption{Normalized Characteristic parameters, GRC and GRG for applied Grey relational analysis to optimize CCD voltages.}
    \label{tab:GRA_result}
\end{table}
\begin{figure}[ht]
\centering
\includegraphics[width=0.5\linewidth]{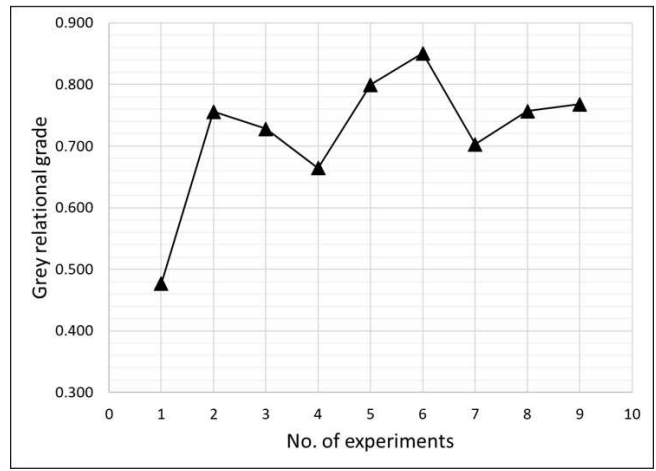}
\caption{Grey relation grade for multiple performance parameters of our CCD-in-CMOS sensor.}
\label{fig:GRA_results}
\end{figure}

\begin{figure}[ht]
\centering
\includegraphics[width=0.7\linewidth]{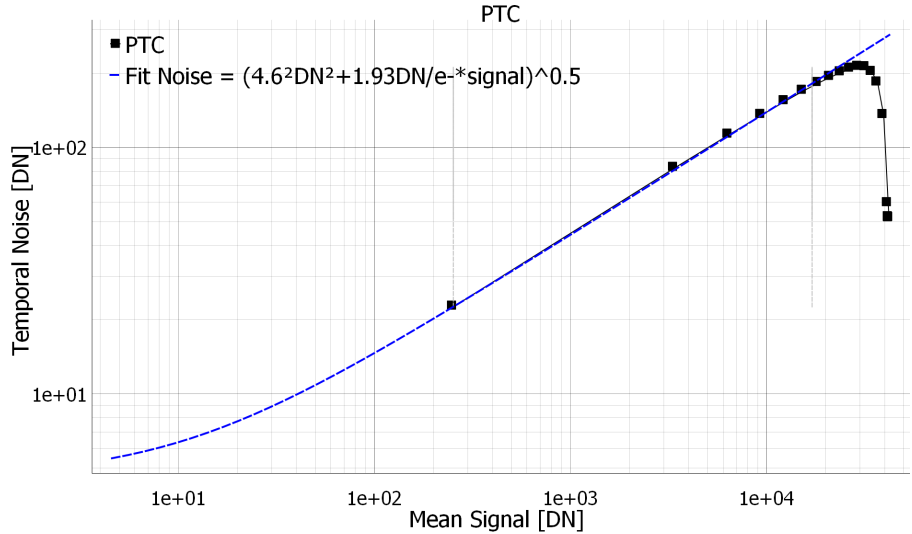}
\caption{Photon transfer curves measured of our CCD-in-CMOS sensor with the optimum voltage setting.}
\label{fig:PTC_results}
\end{figure}

Figure \ref{fig:GRA_results} shows Grey relation grade values from Table \ref{tab:GRA_result}, for all 9 experiments run as per $L_9$ Taguchi orthogonal arrays and it is observed from Figure \ref{fig:GRA_results} also that change in the response when factors go from one level to other. It is also very clear from figure that experiment no 6 has the highest Grey relation grade value. Therefore, CCD voltages of experiment number 6 is likely to be optimal. From Table \ref{tab:Taguchi_result} we can see the voltage setting for experiment number 6 is as follows: $VDDH\_CCD=+1.50V$, $VDDL\_CCD=-1.25V$, $VSSL\_BB=-1.35V$, $VSSL\_OSW=-1.25V$, $VSSL\_OTG1=-0.7V$, and $VSSL\_OTG2=0V$. The photon transfer curve (PTC) with optimum voltage setting have been measured at room temperature as shown in Figure \ref{fig:PTC_results}. A conversion gain of $1.93DN/e^-$ is measured.

\section{Conclusion}
This paper presents a methodology to fine-tune the controlling voltages to reach optimum CCD performance in a large parameter space. This method is applied to our 4-phase CCD-in-CMOS image sensor. We applied Taguchi $L_9$ orthogonal arrays of experiments to reduce the number of experiments with different voltage combinations. Finally, optimal combination of clock voltages is obtained by converting the multiple performance parameters of the detector into a single Grey relational grade. However, we noticed this optimization methodology is more efficient to use for fine-tune the CCD controlling voltages after a preliminary characterization with designed setting. 

\acknowledgments % equivalent to \section*{ACKNOWLEDGMENTS}       
This material is based upon work supported by the funding from the Electronic Component Systems for European Leadership Joint Undertaking under grant agreement No 662222. This Joint Undertaking receives support from the European Union’s Horizon 2020 research and innovation program and Belgium, Netherlands, Greece, France. 

% References
\bibliography{main} % bibliography data in report.bib
\bibliographystyle{spiebib} % makes bibtex use spiebib.bst
\end{document}